\documentclass[journal]{IEEEtran}
\usepackage{graphicx}
\usepackage{flushend}
\begin{document}%
\title{Control of Towing Kites for Seagoing Vessels}%
\author{Michael Erhard and Hans Strauch%
\thanks{Manuscript submitted February 16, 2012; revised July 16, 2012.}%
\thanks{We acknowledge funding from the Federal Ministries {BMWI} and {BMBF},
{LIFE III} of the European Commission, City of Hamburg/BWA and Innovationsstiftung Hamburg.}%
\thanks{M.~Erhard 
is with
SkySails GmbH,
Veritaskai 3,
D-21079 Hamburg,
Germany,
e-mail: michael.erhard@skysails.de,
http://www.skysails.de.}%
\thanks{H.~Strauch is a consultant to SkySails.}
}
\maketitle
\begin{abstract}
  In this paper we present the basic features of the flight control of the
  SkySails towing kite system.
  After introduction of coordinate definitions and basic system
  dynamics we introduce a novel model used for controller design and
  justify its main dynamics with results from system identification based on numerous sea trials.
  We then present the controller design which we successfully use for
  operational flights for several years.
  Finally we explain the generation of dynamical flight patterns.
\end{abstract}
%
\section{Introduction}
\IEEEPARstart{T}{he} SkySails system is a towing kite system
which allows modern cargo ships to use the wind as source of power in order to
save fuel and therefore to save costs and reduce emissions \cite{WWWSkySails}.
The SkySails company has been founded in 2001 and as main business offers 
wind propulsion systems for ships. Starting the development with kites of
6--10\,m$^2$ size the latest product generation with a nominal size of
320\,m$^2$ can replace up to 2\,MW of the main engine's propulsion power. 
Besides the marine applications of kites there is a strongly increasing
activity in using automatically controlled kites
\cite{Canale2007a, Fagiano2010, Williams2008b, WWWKITEnrg,
WWWEnerKite, WWWSwissKitePower} and rigid wings \cite{WWWMakani, WWWAmpyx} in order to
generate power from high-altitude wind \cite{AWEC2011}.
Since 2011 the company's second business segment SkySails Power also develops
and markets systems for generating power from high-altitude wind.
Therefore the design of control systems for tethered kites 
has become a growing field of theoretical
\cite{Ilzhoefer2007, Williams2008a, Williams2008c, Furey2007, Fagiano2009,
Houska2010, Baayen2012} and experimental \cite{Dadd2005, Landsdorp2007a,
Canale2010} research efforts.

\begin{figure} 
  \centering
  \includegraphics[width=8.8cm]{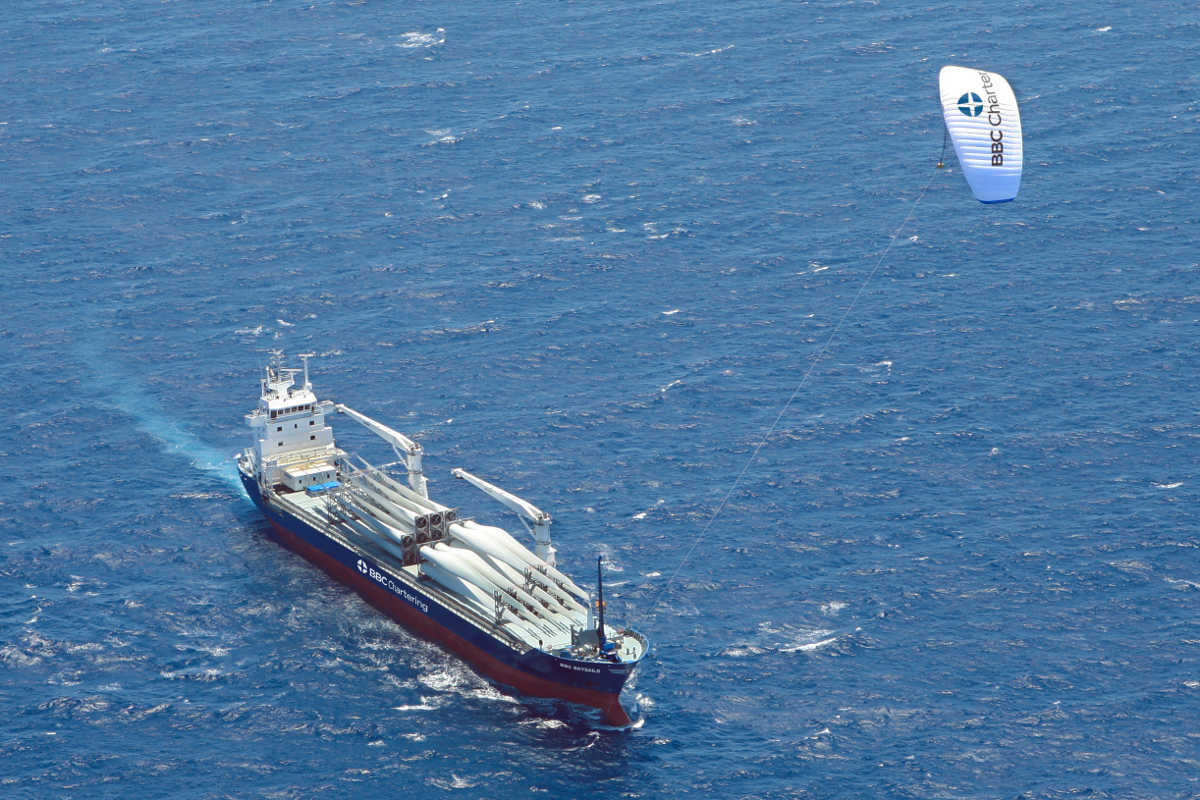}
  \caption{The BBC SkySails with towing kite. The 132\,m vessel utilizes kites
  of sizes up to 320\,m$^2$.}
  \label{fig:sks_system}
\end{figure}

\begin{figure}
  \centering
  \includegraphics[width=8.8cm]{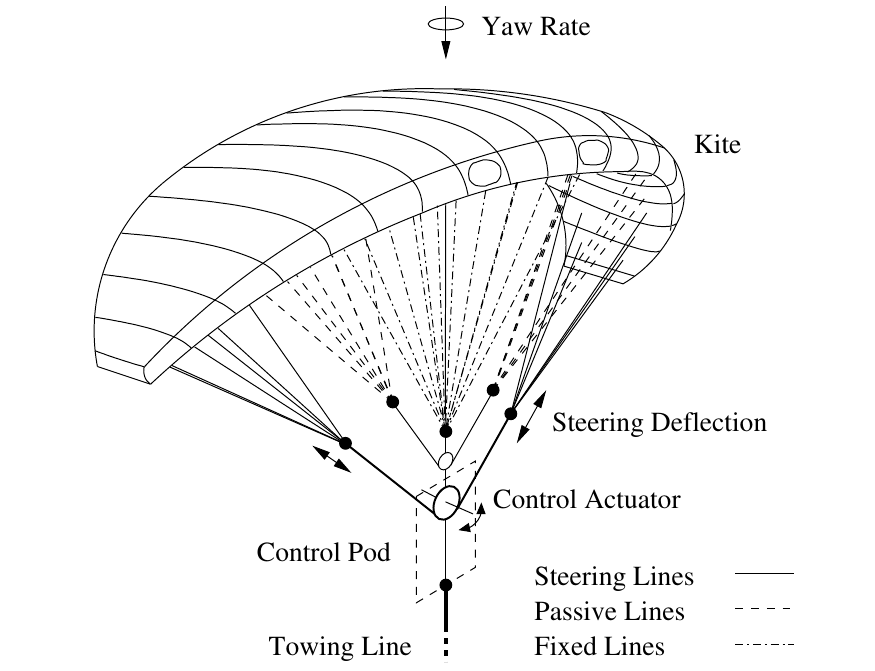}
  \caption{Geometric implementation of the deflection $\delta$ in
  order to direct the kite. The steering actuator in the control pod drives a
  tooth belt attached to the kite steering lines as shown in the figure. The
  main part of the forces is transferred to the control pod by passive and fixed lines. 
  A deflection of the belt warps the kite canopy basically about the roll axis. 
  The resultant dynamics due to aerodynamic forces is mainly a
  turn rate about the yaw axis which is discussed in detail in
  Section \ref{sec:turn_rate_law}.}
  \label{fig:deflection}
\end{figure}

The main components of the SkySails system are shown in
Fig.~\ref{fig:sks_system}
and Fig.~\ref{fig:deflection}.
The core of the propulsion system is the towing kite steered by the control pod
situated under the kite.
The towing force is transmitted to the ship by a high-strength synthetic
fiber rope. Additionally a launch and recovery system is installed aboard the ship
\cite{WWWSkySails}. 
One key component of the flight control system is the main steering
actuator in the control pod applying deflections to some kite lines leading to
curve flight. 

A control system consisting of distributed
computers preprocesses data from various sensors at a rate of 10\,Hz and
performs the flight control algorithm which calculates the steering command
applied to the main actuator. An integrated graphical user interface allows for
operation of the system by the crew whereas for research and development
purposes prototyping and testing toolchains can be connected via special
interfaces.

The paper is organized as follows: First we introduce the basic system
and coordinate definitions. We then focus on the main dynamics and develop a
model specially suited for controller design. After justification of the main
law of the model with experimental data we present our controller design
discussing design considerations and controller performance measurements. We
complete the article with the explanation of pattern generation.
\section{Basic System and Coordinates}
\label{sec:basic_system}
\begin{figure}
   \centering
   \includegraphics[width=8.8cm]{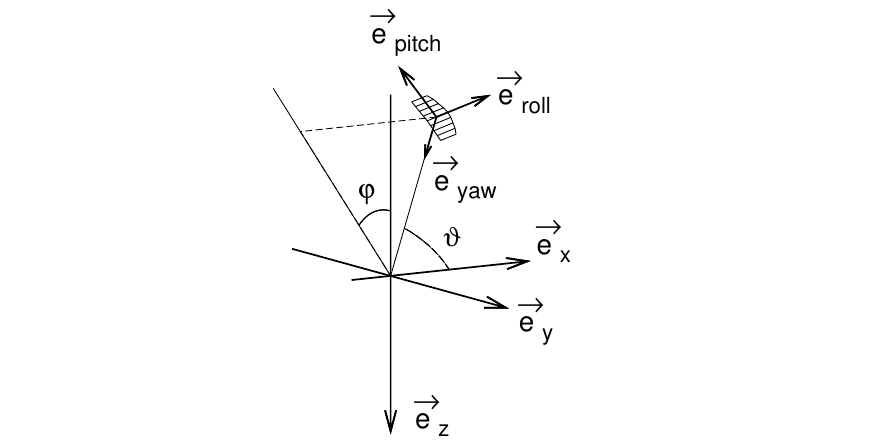}
   \caption{Definition of coordinates for the considered system. 
   The right-handed coordinate system is defined by the basis vectors
   $\vec{e}_x, \vec{e}_y, \vec{e}_z$ with $\vec{e}_x$ in wind direction and
   $\vec{e}_z$ pointing downwards with respect to gravity. 
   The kite position is parameterized by introducing the spherical coordinates
   $\varphi$ and $\vartheta$ (for a more precise definition see
   (\ref{eq:def_rotation})).
   The kite axes are labeled as roll $\vec{e}_{\rm roll}$, pitch $\vec{e}_{\rm
   pitch}$ and yaw $\vec{e}_{\rm yaw}$. This corresponds to the definition usually used
   in aerospace applications with roll axis parallel to forward and yaw axis
   parallel to down directions respectively.
   Note that the yaw vector $\vec{e}_{\rm yaw}$ is defined by the position
   of the kite assuming it is constrained to the origin by a rigid rod.
   Thus orientation of the kite is represented by the single angle $\psi$.
   Detailed vector definitions are given in Appendix \ref{sec:derivation_model}
   of the paper.}
   \label{fig:coordinate_definition}
\end{figure}

In this section we give a mathematical description of the considered system.
It is worth mentioning that we deal with a constrained system which shows
a completely different dynamics compared to free flying parafoils \cite{JLingard1986}.
The basic system is sketched in Fig.~\ref{fig:coordinate_definition}.

Compared to the real system we make use of some simplifications which are
summarized in Table \ref{tab:simplifications}.
\begin{table}[b!]
  \renewcommand{\arraystretch}{1.3}
  \caption{Overview on Model Assumptions used for Setup  
  (Section \ref{sec:basic_system}) and Dynamics
  (Section \ref{sec:system_dynamics})}
  \label{tab:simplifications}
  \centering
  \begin{tabular}{l|p{6cm}}
     \hline\hline
     Masses neglected & 
     Usually the aerodynamic forces are larger than system
     masses and thus acceleration effects play a minor role. The system can
     be considered to be in equilibrium flight state.
     This assumption simplifies the equations of motion significantly.\\
     %
     %
     \hline
     \begin{minipage}[t]{1.8cm}\flushleft Rope dynamic neglected\end{minipage}
     &
     Apart from exceptional situations the towing rope acts as rigid tether
     and is considered as massless tether only. Winching during launch and
     recovery is not considered in this paper.\\
     %
     \hline
     Aerodynamics &
     The aerodynamics of the kite is reduced to two assumptions. First we assume
     the kite is always in its aerodynamic equilibrium which means that the air
     flow is determined by the glide ratio $E$ (compare Appendix
     \ref{sec:derivation_model}). Secondly the response to a steering deflection 
     can be described mainly by one parameter $g$ as we show by experimental
     data in Section \ref{sec:turn_rate_law}.\\
     %
     \hline
     Wind field & We assume a constant and homogeneous wind field with velocity
     $v_0$ to derive the equations. As this assumption often does not
     hold for real situations we either use the average wind speed at flight altitude, which has to be
     estimated, for $v_0$ or --- as done for the controller setup --- 
     the air path speed $v_{\rm a}$ (see Appendix \ref{sec:eqm_summary}).\\
     %
     \hline
     \begin{minipage}[t]{1.8cm}\flushleft Vessel dynamic
     neglected\end{minipage} &
     As forward force optimization is not treated in this paper, course and
     speed of the vessel can be easily eliminated by considering them in the relative
     wind speed and direction.\\ 
     \hline\hline
  \end{tabular}
\end{table}
The flexible rope is substituted by a rigid rod which also is parallel to the
kite yaw axis $\vec{e}_{\rm yaw}$ and all masses are neglected.
At first view this seems to be an unrealistic simplification,
but the usual mode of operation is the highly dynamical pattern flight leading to
line forces large compared to system masses. Therefore inertia effects or free
flight situations, where the towing line is no longer stretched out due to gusts
or wave induced ship motions, are infrequent. Although consideration of these issues becomes important at a certain point when bringing the system to higher perfection, a detailed
description of the solution to these off-nominal situations would go beyond the scope of this paper.

\begin{figure}
   \centering
   \includegraphics[width=8.8cm]{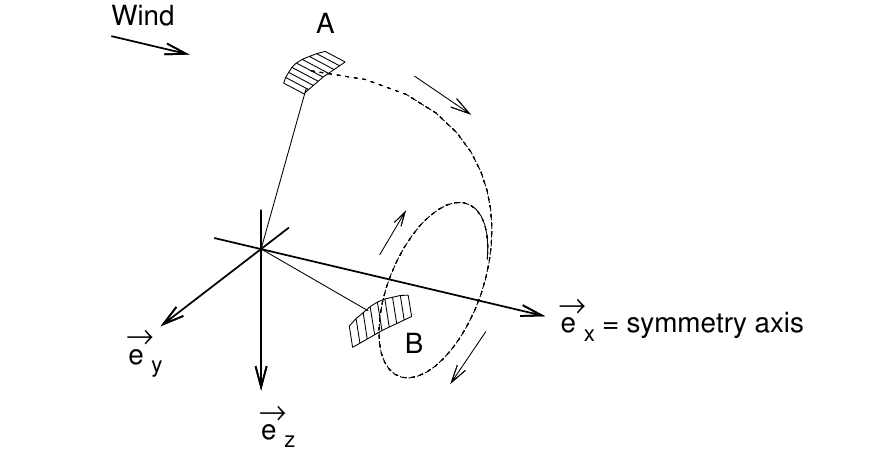}
   \caption{Flying a kite in a fictitious wind tunnel on a space station would
   be instructive in understanding the basic equations. While the neutral flight
   is stationary at any arbitrary position like for example A, a constant
   $\psi\neq 0$ due to steering will lead to a circular orbit B. The
   diameter is a function of the $\psi$ magnitude. See also
   Fig.~\ref{fig:windwindow} for further illustration.}
   \label{fig:space_exp}
\end{figure}

We would like to start with a demonstrative and introductive example to
motivate the idea of the chosen coordinate system $\varphi, \vartheta$ and to
explain the basic dynamics. Imagine a kite flying in a wind tunnel experiment
conducted on a space station which means the absence of gravity. We further assume the
free manoeuvrability of the kite unrestricted by obstacles like ground or ceiling.
From this mental picture it would be natural to chose a coordinate system with symmetry axis
in wind direction.
A kite with its roll axis $\vec{e}_{\rm roll}$ antiparallel to the wind
direction would stay at a some arbitrary, stationary position like e.g. 'A', see
Fig.~\ref{fig:space_exp}. 
However, once a deflection is
commanded, the kite will go 'down' and orbit around permanently on a circular
path 'B'. The diameter of the steady state circle is a function of our
coordinate $\psi$ which in turn is determined by the commanded deflections.
The smaller the diameter of the circle the faster flies the kite and thus the
more force will be generated. The corresponding dynamical equations will be
discussed in detail in Section \ref{sec:system_dynamics}.

We would like to emphasize that the above sketched model is our controller
design model while we use for various other development and test issues  
a sophisticated simulator model including multi-body dynamics which
also captures the aerodynamical effects more
comprehensively (comparable in structure to \cite{PWillams2007}). Yet, we
suggest above model based on this imaginary space station experiment because it
is specifically suited for the controller design purpose. It neglects gravity
effects and by this a new symmetry axis is gained. The main benefit of this
symmetry compared to other coordinate systems
\cite{Williams2008b,Diehl2001a,Dadd2011} is the resulting simplicity of
equations of motion which allows us to approach the feedback and guidance design
task to a large extent analytically or semi-analytically at least.
Further on it provides a straight forward way of describing flight patterns.

Our choice of the design model led to a controller structure to be presented in
sections
\ref{sec:controller_design_overview}--\ref{sec:controller_design_outer_loop}.
This controller structure turned out to work quite effectively in numerous sea trials. 

We close this section by definitions for the quantities used in the
following. The coordinate system is shown in figure
\ref{fig:coordinate_definition}.
For a constant line length $L$ the state of the kite is defined by the three
angles $\varphi, \vartheta$ and $\psi$. With respect to the basis vectors
$\vec{e}_x$, $\vec{e}_y$, $\vec{e}_z$ we obtain for the kite position $\vec{x}$:
\begin{equation}
    \vec{x} = L
    \left( \begin{array}{c}
    \cos\vartheta \\
    \sin\varphi \sin\vartheta \\
    -\cos{\varphi}\sin\vartheta
    \end{array}\right).
\end{equation}

The kite axes are denoted as $\vec{e}_{\rm roll}$ (roll or longitudinal),
$\vec{e}_{\rm pitch}$ (pitch) and $\vec{e}_{\rm yaw}$ (yaw). An explicit definition of
these vectors is given in Appendix \ref{sec:derivation_model}.

For a description using rotation matrices one would start with a kite at
position $L\vec{e}_x$ with roll-axis in negative $z$-direction
$\vec{e}_{\rm roll}=-\vec{e}_z$ and then apply the following rotations: $-\psi$
about $x$, $\vartheta$ about $y$ and finally $\varphi$ about $x$.
This transformation reads:
\begin{equation}
  R=R_x(\varphi)R_y(\vartheta)R_x(-\psi). \label{eq:def_rotation}
\end{equation}

One could interpret the angle $\psi$ as orientation of the kite
longitudinal axis with reference to the wind. For a given kite
position $\vec{x}$ (parameterized by $\varphi$ and $\vartheta$) the reference
orientation $\psi=0$ corresponds to the minimum of the scalar product
$\left(\vec{e}_{\rm roll}, \vec{e}_x\right)$ obtained when turning the kite fixated at
this position $\vec{x}$ around its yaw axis $\vec{e}_{\rm yaw}$.
A nonzero value $\psi$ represents a kite orientation obtained by a rotation of
$\psi$ about the yaw axis $\vec{e}_{\rm yaw}$ starting at this reference orientation.
\section{System Dynamics used for Design}
\label{sec:system_dynamics}
For verification and other development purposes we use a full dynamics
simulation containing a multi-body model, an aerodynamic database and
parameters adopted to results of sea trials. In this section we would
like to present the main relations and the dynamics of a complementary model
specifically tailored for the design of the controller. The detailed derivation
steps are summarized in Appendix \ref{sec:derivation_model}.
\begin{figure}
  \centering
  \includegraphics[width=8.8cm]{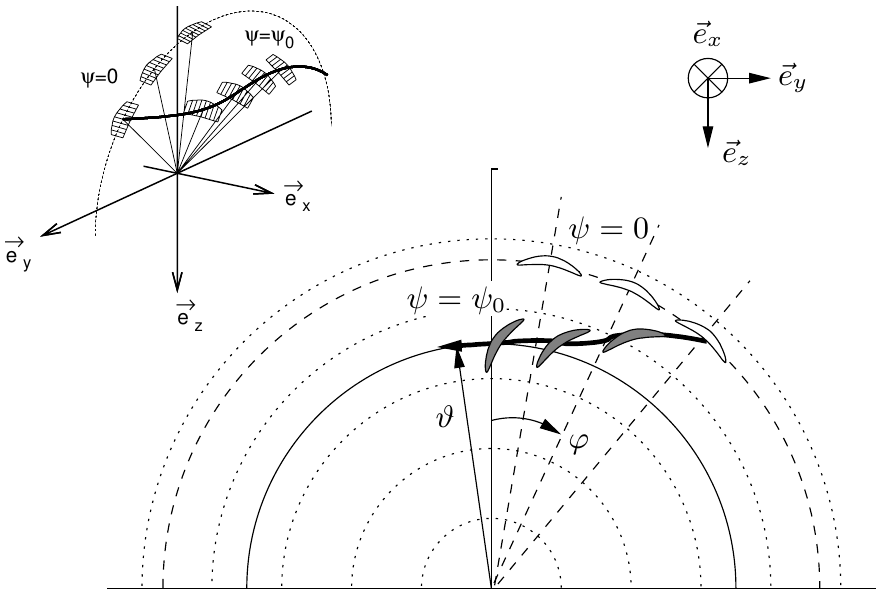}
  \caption{Zenith positions for neutral flight with
  low force.
  With $\psi =0$ 
  the parameter $\varphi$ can be freely chosen determining
  the positions (drawn as white kites). For constant $\psi$ the resulting
  $\vartheta$ converges as shown by the gray kites flying on the marked
  trajectory.}
  \label{fig:windwindow}
\end{figure}
The equations of motion for $\vartheta$ and $\varphi$ read:
\begin{eqnarray}\label{eq:kinematic1}
  \dot{\vartheta} &=& \frac{v_{\rm a}}{L} \left(
  \cos\psi - \frac{\tan\vartheta}{E}\right) \label{eq:dot_theta}\\
  \dot{\varphi} &=& -\frac{v_{\rm a}}{L \sin\vartheta} \sin\psi.\label{eq:dot_phi}
\end{eqnarray}
Thus the dynamic is mainly controlled by the angle $\psi$. Further quantities
are the air path speed $v_{\rm a}$, the towing line length $L$ and the glide ratio
$E$. As already pointed out and to be reasoned in the next section we neglect
acceleration and gravity effects in order to obtain these simple equations of
motion which allow for the following interpretations.

Detailed computation steps for the subsequent two steady state solutions are
given in Appendix \ref{sec:derivation_model}. First, for constant $\psi$, we get
a flight trajectory on a circle with constant angle $\vartheta_0$ given by
\begin{equation}
  \vartheta_0(\psi) = \arctan(E \cos \psi).\label{eq:vartheta_psi}
\end{equation}
This equation also applies to the fictitious space station
experiment introduced in Section \ref{sec:basic_system} and Fig.~\ref{fig:space_exp}. 
A further result is the dependence of the (steady state) air path velocity $v_{\rm a}$
on the value of $\psi$ and the ambient wind speed $v_0$,
\begin{equation}
  v_{\rm a} = v_0  E \cos\vartheta = v_0  E \cos(\arctan (E \cos\psi))
  \label{eq:v_a_theta}
\end{equation}
which is the key issue for pattern generation as we can use $\psi$ as a
tuning knob to control $v_{\rm a}$ and the force which is proportional to $v_{\rm a}^2$
accordingly.
It is worth mentioning that for practicing the sport of kite surfing, the
content of (\ref{eq:v_a_theta}) is crucial in order to control forces: kite
surfers know the angle $\vartheta$ as position in the so called 'wind window' and the deeper they fly
their kite into this 'wind window' --- i.e.~they decrease this angle --- the
more traction force they get and vice versa.

A special case is the so called zenith position for $\psi=0$, a neutral flight
situation which generates only low forces and thus is used mainly for launching
and recovering the kite. From (\ref{eq:vartheta_psi}) we get $\vartheta_0 =
\arctan(E)$ and $\varphi$ can be used as free parameter to determine the neutral
flight position (compare Fig.~\ref{fig:windwindow}).

While all equations up to here follow straightforward from
kinematic reasoning, albeit the choice of the coordinate system was not
'typical', the dynamic response of the kite to a steering deflection $\delta$ is
claimed to be 
\begin{equation}
   \dot{\psi}_{\rm m} = g\, v_{\rm a}\, \delta \label{eq:Drehratengesetz}
\end{equation}
where $g$ is the proportionality factor and an illustration of the deflection
$\delta$ is given in Fig.~\ref{fig:deflection}.

We would like to draw special
attention to this turn rate law (\ref{eq:Drehratengesetz}) and will show in the following that it can
be justified by measured data to a surprising high degree.
Therefore it is a key issue for the cascaded controller approach where it
constitutes the dynamics of the inner loop.

Finally we would like to point out that due to its motion on a spherical surface
an inertial sensor measures a turn rate $\dot{\psi}_{\rm m}$ about the yaw axis
$\vec{e}_{\rm yaw}$ different from the derivation $\dot{\psi}=d\psi/dt$.
The rotation measured by the pod sensors can be calculated by
transforming the dynamics represented by $\dot{R}$ into the pod coordinate
system by applying (\ref{eq:def_rotation}). Comparing the rotation operation
$\vec{\Omega}\times$ with $R\!\cdot\!\dot{R}^T$ yields:
\begin{equation}
  \dot{\psi}_{\rm m} = \dot{\psi} - \dot{\varphi} \cos\vartheta.
\end{equation}
By consideration of typical flight situations where either $\varphi$ is
kept constant during the neutral flight mode or $\dot{\varphi} \propto v_{\rm a}/L$ 
becomes small due to the long line length $L$ needed for dynamic pattern
flight, one can assure oneself that the second term of this equation usually is
small compared to the first and thus can be neglected in the first instance and
treated as a correction to the controller design later.

We would like to conclude this section by emphasizing that we
presented a novel model based on three state variables $\psi$, $\vartheta$,
$\varphi$ and three equations of motion (\ref{eq:Drehratengesetz}),
(\ref{eq:dot_theta}), (\ref{eq:dot_phi}) whereas previously published models \cite{Canale2007a}, \cite{Fagiano2010},
\cite{Ilzhoefer2007}, \cite{Houska2010} introduce at least four or more
state variables.
For a summary and further discussion of the equations we refer to Appendix \ref{sec:eqm_summary}.
\section{Justification of the Turn Rate Law}
\label{sec:turn_rate_law}
In this section a justification of (\ref{eq:Drehratengesetz}) is given based on
numerous experiments showing the strong proportionality. It is worth
noting that (\ref{eq:Drehratengesetz}) is confirmed by measurements to a
high degree even in disturbed sea trial conditions.
The key issue of these experiments is to perform
bang-bang flights which will be presented and discussed in the following.

The excitation of the system for the identification is performed in the following
way: We apply a constant steering command
$+\delta_0$ to the system. The system will respond with a positive yaw rate
($\dot{\psi}_{\rm m}\!>\!0$). When reaching a certain threshold $\psi_{\rm m}\!\geq\!
\psi_0$ the corresponding opposite steering deflection $-\delta_0$ is commanded
leading to a decrease of $\psi_{\rm m}$. Falling below the negative threshold
$\psi_{\rm m}\!\leq-\!\psi_0$, the primary deflection $\delta_0$ is commanded again.
The schedule of the bang-bang experiments is as follows: the human pilot flies
the kite into a high zenith position with $\varphi\!\approx\! 0$ and hands over
the steering to the computer based control system which performs the described algorithm.

\begin{figure}
  \centering
  \includegraphics[width=8.8cm]{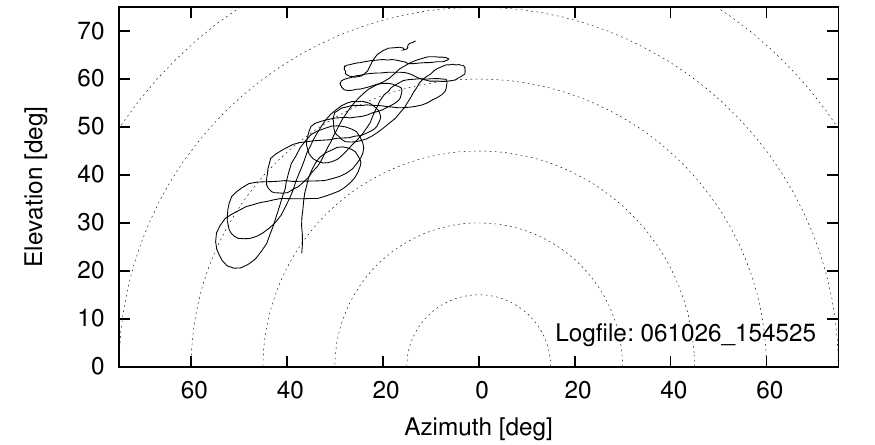}
  \caption{Flight trajectory under computer control
  for the bang-bang flight experiment. Angles were measured by tow point sensors
  on the ship determining the direction of the towing rope.}
  \label{fig:sysident1}
\end{figure}

A typical flight trajectory is shown in Fig.~\ref{fig:sysident1}.
The bang-bang steering leads to a figure-eight-pattern and for
typical parameters the air path speed and thus the size of the
pattern increase because the kite flies down to smaller elevation angles
$\vartheta$.
The human pilot only has the task to supervise the flight and overtake manual
control before the system runs into the danger of overload or bounces
against the water surface.

\begin{figure}
  \centering
  \includegraphics[width=8.8cm]{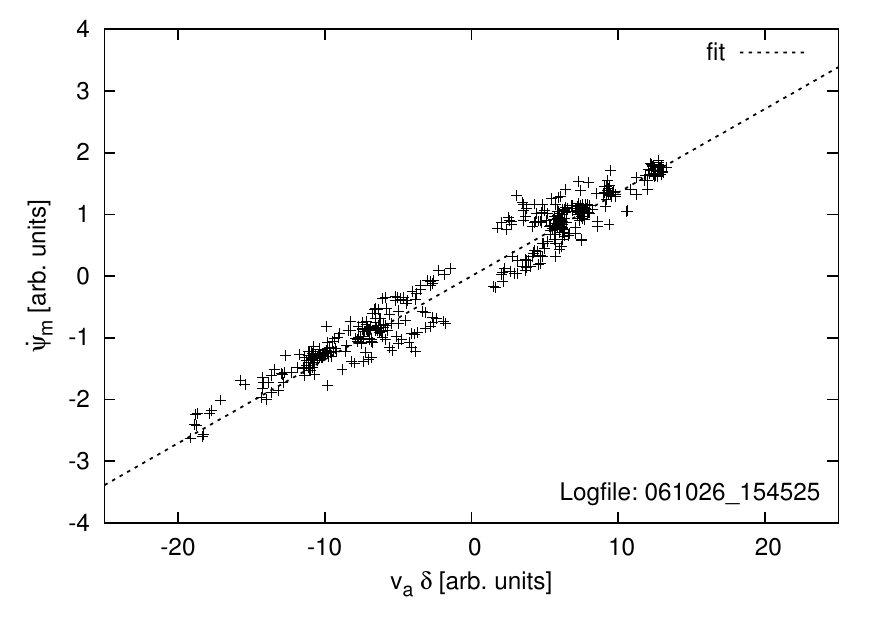}
  \caption{Measured data of a bang-bang flight.
  Yaw rate $\dot{\psi}_{\rm m}$ as function of the air path
  speed multiplied by deflection $v_{\rm a} \delta$ to justify
  (\ref{eq:Drehratengesetz}).
  The parameter $g$ is obtained by the shown linear fit.}
  \label{fig:sysident2b}
\end{figure}

In Fig.~\ref{fig:sysident2b} data points for one experiment run are shown. This
experiment was performed in 2006 using a 20\,m$^2$ kite.
Repeating this experiment with different $\delta_0$ leads to similar
$g$ values and thus proves the validity of (\ref{eq:Drehratengesetz}).

\begin{figure}
  \centering
  \includegraphics[width=8.8cm]{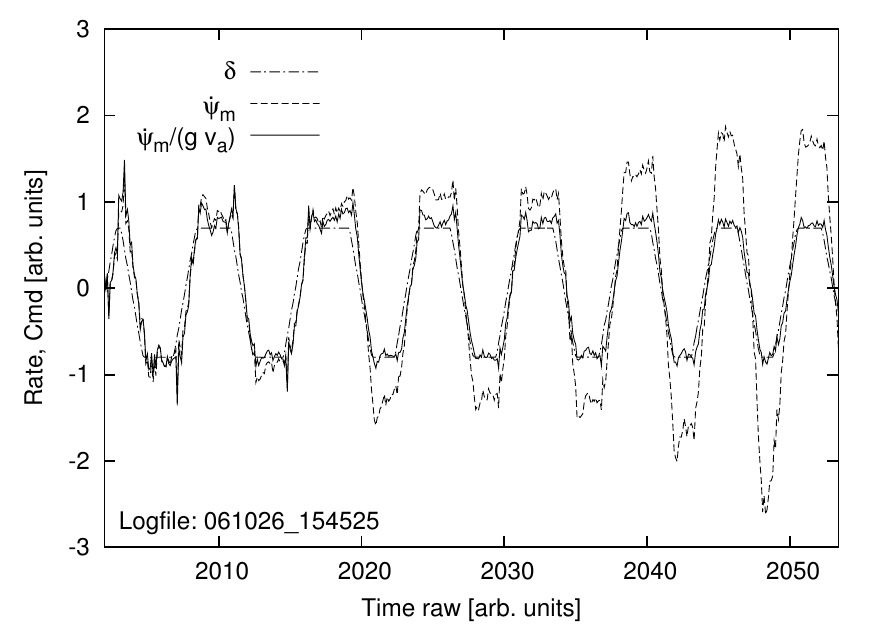}
  \caption{Comparison of steering command $\delta$ with yaw rate
  $\dot{\psi}_{\rm m}$ and the yaw rate divided by the air path velocity
  $\dot{\psi}_{\rm m}/(g\,v_{\rm a})$. Note the increasing rates are due to increasing
  air path speed $v_{\rm a}$ while going down into the wind window (compare
  Fig.~\ref{fig:sysident1}).}
  \label{fig:sysident3}
\end{figure}

Although the linear dependence can be clearly identified in
Fig.~\ref{fig:sysident2b} it is even more convincing to present the data
in the time domain as shown in Fig.~\ref{fig:sysident3}.
Here we compare the time-series of the steering command with the turn rate of
the kite. The trapezoidal shape  is due to the finite steering
velocity of the control pod. The resulting measured yaw rate $\dot{\psi}_{\rm m}$
shows an increase which results from the increasing air path velocity $v_{\rm a}$ over the
experiment. The yaw rate divided by $g v_{\rm a}$ is also plotted in order to compare
it to the steering command. Although a lot of perturbations affect these experiments we
observe an excellent correlation. This analysis justifies the
validity of (\ref{eq:Drehratengesetz}) to a high degree and recommends its usage
as a key role for the controller design.

At this point we would like to classify and review these bang-bang experiments
in a historical context. Following the textbook approach in
classical system identification we did a lot of identification flights in the years
2005 and 2006 using separate batch runs in order to characterize the steering behavior of our kites
at the various operating points.
The experiments were quite cumbersome: perturbations from wind gusts can be
comparatively large compared to periodic excitations caused by the deflection commands
and the air path velocities
were difficult to tune for the different batch runs and desired operating points. 
As we had to evaluate data
from different days with changing environmental conditions and drifting flight
properties of our kites we solely could suspect the validity of law
(\ref{eq:Drehratengesetz}).
But once we switched to a bang-bang flight strategy the real law shows
up clearly. This is obvious as one bang-bang experiment varies the parameter
$v_{\rm a}$ over a whole range, in one flight alone, lasting only a few tens of seconds
and therefore plays a trick on perturbations. 

We would like to conclude this section by giving an extended version of
(\ref{eq:Drehratengesetz}) which also takes into account the effect
of the gravitational force on the turn rate and reads:
\begin{equation}
  \dot{\psi}_{\rm m} = g\, v_{\rm a}\, \delta + M
  \frac{\cos\theta_{\rm g} \sin\psi_{\rm g}}{v_{\rm a}}. \label{eq:Drehratengesetz_with_mass_term}
\end{equation}
The quantity $\theta_{\rm g}$ denotes the angle between $\vec{e}_{\rm yaw}$ and the
$\vec{e}_x$-$\vec{e}_y$-plane and $\psi_{\rm g}$ the angle between $\vec{e}_{\rm
roll}$ and the $\vec{e}_x$-$\vec{e}_y$-plane.

Because a steering deflection could be regarded as a kite force component into pitch
direction $\vec{e}_{\rm pitch}$ subsequently leading to a yaw rate, the gravity force,
projected onto the pitch axis by $\cos\theta_{\rm g} \sin\psi_{\rm g}$, should have
the same effect.
We have shown in this section that the yaw rate is proportional to
$v_{\rm a}\,\delta$. This can be attributed to a side force proportional to
$v_{\rm a}^2\,\delta$ from aerodynamical and design considerations.
Transferring this reasoning to the mass term, which is independent from $v_{\rm a}$, we
expect a factor of $M/v_{\rm a}$ between the 'gravitational' side force and the yaw
rate. The constant $M$ includes system masses and kite characteristics.
As $M$ is positive for our kites we get an instable behaviour and thus have to
stabilize $\psi_{\rm m}$ by active control.

As for the usual operation point of dynamic flight we have $(g v_{\rm a}) \gg (M/v_{\rm a})$
and thus the second term of (\ref{eq:Drehratengesetz_with_mass_term}), which we
call 'mass term', can be neglected for the design of the linear feedback law only
to be introduced as correction term via a feedforward path to the controller structure.

As our operational flights under autopilot conditions (during the dynamic flight modes) utilize similar bang-bang like commands, we can use the discussed identification scheme as a standard tool
to establish or check the kite parameter during normal operation.
During flights a recursive least-square algorithm
\cite{Ljung2007a} runs in order to determine the system parameters $g$ and $M$
on-line. We monitor these values for changes which may indicate upcoming material failures in advance and can
adapt controller parameters on-line, if necessary, to increase robustness.

After we have convinced ourselves of the validity of (\ref{eq:Drehratengesetz}),
we will now  present the controller design which is strongly influenced in its structure by the discussed law.
\section{Controller Design}
\label{sec:controller_design_overview}
Most of time the kite is operated in a highly dynamic regime where the
air path speed can easily vary by a factor of up to 3--5 within some seconds
and the deflection command can change by more than $60 \%$.
The classical way to approach such a controller design task would be to use a controller 
structure which specifically aims at time varying, non-linear systems. Non-linear dynamic
inversion or non-linear model predictive control could be such candidates. Gain scheduling based 
on linearized plant models along the trajectory is a further alternative.
Actually we tried the latter approach in the beginning but were not satisfied
with the achievable robustness. The core of the problem is governed by
the fact that it is difficult to execute a classical modeling approach as
usually performed in aerospace application. Such an
approach would be based on an aerodynamic database covering the full dynamic
regime.
Performing wind tunnel tests would be expensive and for our larger kites with
160\,m$^2$ area downright impossible. We
also learned that a controller structure where the kite trajectory is given as a
set-point directly into a single controller block, which then directly
computes the deflection command, lacks robustness due to the above mentioned modeling issue. 

\begin{figure}
  \centering
  \includegraphics[width=8.8cm]{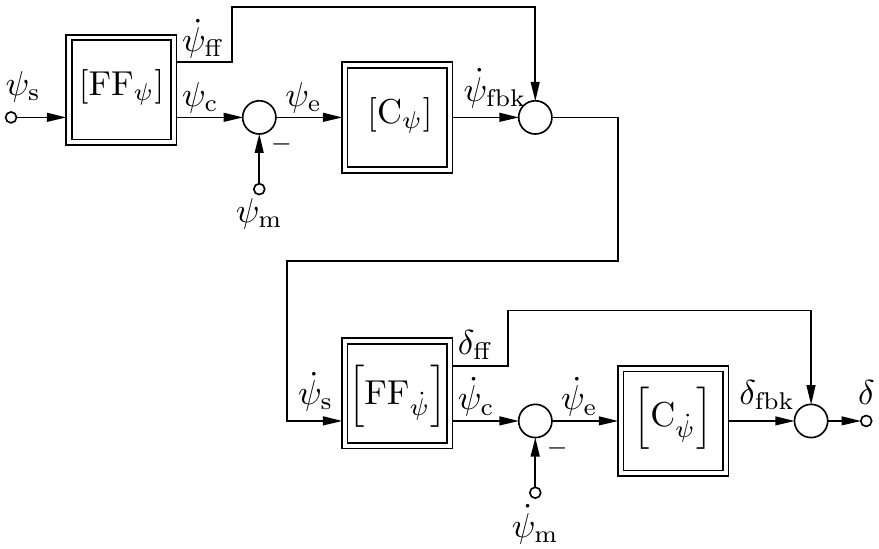}
  \caption{Cascaded controller approach for $\psi$ control implementing the model following 
  structure. Detailed diagrams
  for the blocks $[{\rm FF}_{\psi}]$, $[{\rm C}_\psi]$, $[{\rm FF}_{\dot{\psi}}]$ and
  $[{\rm C}_{\dot{\psi}}]$ are shown in figs.~\ref{fig:block_ff_psi},
  \ref{fig:block_c_psi},
  \ref{fig:block_ff_psidot} and \ref{fig:block_c_psidot}. The Controller
  calculates a steering command $\delta$ from the input value $\psi_{\rm s}$ 
  using a measured yaw angle $\psi_{\rm m}$ and yaw rate $\dot{\psi}_{\rm m}$.}
  \label{fig:cascaded_controller}
\end{figure}

Instead we settled for a separation
of the overall controller task into 'guidance' and 'control'. Section \ref{sec:patterngeneration} describes
how the guidance algorithm computes $\psi_{\rm s}$, which is then the input to the
controller. However the major distinguishing feature of our autopilot, compared
to other approaches we found in the cited literature, is the cascaded controller structure which
is based on the model following principle as shown in Fig.~\ref{fig:cascaded_controller}.

The basic idea is to reflect the separation of dynamics
(\ref{eq:Drehratengesetz}) and kinematics, as given by $\dot{\psi}=d\psi/dt$,
adequately in the controller structure.
We can see two cascaded loops in Fig.~\ref{fig:cascaded_controller}. The
inner loop gets a commanded rate $\dot{\psi}_{\rm s}$ as an input and computes the
deflection command $\delta$. The outer loop has ${\psi}_{\rm s}$ as input and
commands $\dot{\psi}_{\rm s}$ to the inner loop.
Before discussing each controller element in detail we summarize all variables
used and relate them to measurements in Table \ref{tab:variables}.
\begin{table}[b!]
    \renewcommand{\arraystretch}{1.3}
  \caption{Variables for Conroller Design}
  \label{tab:variables}
  \centering
    \newcommand{\tableWidth}{7cm}
  %
  \begin{tabular}{l|p{\tableWidth}}
  
     %
     %
     \multicolumn{2}{l}{Control Actuation} \\
     \hline\hline
       $\delta$ & Normalized steering deflection (see Fig.~\ref{fig:deflection})\\
       \quad $\delta_{\rm ff}$ & \quad Feedforward computed as shown in
       Fig.~\ref{fig:block_ff_psidot}.\\
       \quad $\delta_{\rm fbk}$ & \quad Feedback computed as shown in
       Fig.~\ref{fig:block_c_psidot}.\\
     \hline 
     \multicolumn{2}{c}{}\\
     %
     %
     %
     \multicolumn{2}{l}{System Dynamics} \\
     \hline\hline
       %
       $\psi$ & Orientation angle of kite longitudinal axis $\vec{e}_{\rm
       roll}$ with respect to the ambient wind (see Section \ref{sec:basic_system}).
       \\
       \quad $\psi_{\rm s}$ & \quad Setpoint value from guidance (pattern
       generation)\\
       \quad $\psi_{\rm c}$ & \quad Control reference computed as shown
       in Fig.~\ref{fig:block_ff_psi}.\\
       \quad $\psi_{\rm m}$ & \quad \parbox[t]{6.5cm}{Value based
       on inertial measurement unit located in the control pod (pod-IMU) 
       and wind direction estimate.}
        \\
       \quad $\psi_{\rm e}$ & \quad Control error ($\psi_{\rm e}\!=\! \psi_{\rm
       m}\!-\!\psi_{\rm c}$)\\
     \hline
       %
       $\dot{\psi}$ & Turn rate about yaw axis \\
       \quad $\dot{\psi}_{\rm ff}$ & \quad Feedforward computed as shown in
       Fig.~\ref{fig:block_ff_psi}.
       \\
       \quad $\dot{\psi}_{\rm fbk}$ & \quad Feedback computed as shown in
       Fig.~\ref{fig:block_c_psi}. \\
       \quad $\dot{\psi}_{\rm s}$ & \quad Setpoint value
       $\dot{\psi}_{\rm s}\!=\!\dot{\psi}_{\rm ff}\!+\!\dot{\psi}_{\rm fbk}$.\\
       \quad $\dot{\psi}_{\rm c}$ & \quad Control reference computed as
       shown in Fig.~\ref{fig:block_ff_psidot}.\\
       \quad $\dot{\psi}_{\rm m}$ & \quad Measured value based on pod-IMU  \\
       \quad $\dot{\psi}_{\rm e}$ &
       \quad Control error 
       ($\dot{\psi}_{\rm e}\! =\! \dot{\psi}_{\rm m}\!-\!\dot{\psi}_{\rm c}$)\\
     \hline
       %
       $\theta_{\rm g}$ & Angle between yaw axis and
       $\vec{e}_x$-$\vec{e}_y$-plane, based on pod-IMU measurement.\\
     \hline
       $\psi_{\rm g}$ & Angle between roll axis and
       $\vec{e}_x$-$\vec{e}_y$-plane, based on pod-IMU measurement.\\
     \hline
       $v_{\rm a}$ & Airpath speed as measured with respect to $\vec{e}_{\rm
       roll}$ by an anemometer located aboard the control pod.\\
       \quad $K_{\dot{\psi}}$ & \quad \parbox[t]{6.5cm}{Current
       gain ($K_{\dot{\psi}}\!=\!g v_{\rm a}$) between turn rate and deflection
       $\dot{\psi}=K_{\dot{\psi}}\delta$ (see
       Section \ref{sec:controller_design_inner_loop})}\\
       \quad $T_1$ & \quad Influence of gravity on yaw rate, see (\ref{eq:T_1}).
       \\
     \hline 
       $\varphi$ & Wind window position, see Fig.~\ref{fig:windwindow}\\
       \quad $\varphi_{\rm m}$ & \quad 
       \parbox[t]{6.5cm}{Measured
       value based on wind direction estimate and angular sensors at ship
       towpoint which are wave-motion compensated by the
       ship-IMU.\vspace{2pt}}\\
     \hline
       $\vartheta$ & Wind window position, see Fig.~\ref{fig:windwindow}.\\
     \hline
     \multicolumn{2}{c}{}\\
     \multicolumn{2}{l}{System Parameters}\\ 
     \hline\hline
       $g$ & Proportional gain of the turn rate law (\ref{eq:Drehratengesetz}),
       for a 160\,m$^2$ system we find $g\approx 0.03$--0.05\,rad/m.\\
       \hline
       $M$ & Effect of gravitation on turn rate, see
       (\ref{eq:Drehratengesetz_with_mass_term})\\
       \hline
       $E$ & Glide ratio $L/D$, typically $E=$4--5 in our case.\\
       \hline
       $L$ & Tether line length assumed to be constant as launch and
       recovery are not considered here.\\
       \hline
       $\dot{\delta}_{\rm p}$ & Steering speed of the control pod (typically
       0.3--0.5\,1/s)\\
       \hline
       $v_0$ & Ambient wind speed defined for model.\\
      \hline
  \end{tabular}
\end{table}
\section{Controller Inner Loop}
\label{sec:controller_design_inner_loop}
From (\ref{eq:Drehratengesetz}) we recognize that the dynamics from deflection to yaw rate
can be viewed as a proportional plant, $\dot{\psi}_{\rm m}=K_{\dot{\psi}}\,\delta$,
where  $K_{\dot{\psi}}=g\,v_{\rm a}$ denotes the gain. Of course we have to realize
that $K_{\dot{\psi}}$ is not constant but a function of the air path velocity.
We take care of this by employing a feedforward/feedback structure which implements
the model following principle:
in the feedforward term $[{\rm FF}_{\dot{\psi}}]$ we compute, in an open loop
fashion, the deflection command $\delta_{\rm ff}$ necessary to achieve 
$\dot{\psi}_{\rm s}$ .
The feedback control $[{\rm C}_{\dot{\psi}}]$ only acts when there is a
remaining control error due to external disturbances or due to unmodeled plant dynamics. An appropriate delay block is necessary in order to capture all the delays from command to actual execution.

\begin{figure}
  \centering
  \includegraphics[width=8.8cm]{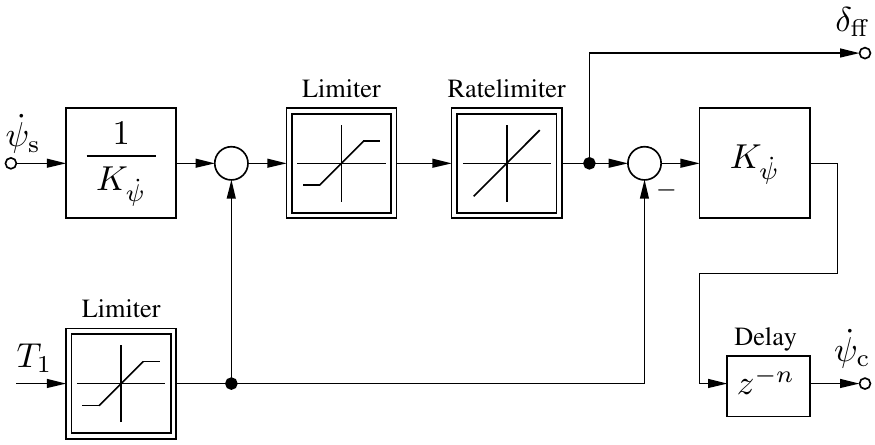}
  \caption{Details of block $[{\rm FF}_{\dot{\psi}}]$ of the cascaded controller
  (see Fig.~\ref{fig:cascaded_controller}). From input value $\dot{\psi}_{\rm s}$ the
  feedforward value $\delta_{\rm ff}$ and $\dot{\psi}_{\rm c}$ are calculated
  by using a steering pod model consisting of a limiter and a ratelimiter. Note
  the translation of rate to command and back via division and
  multiplication by $K_{\dot{\psi}}$. Various delays in the whole loop   are
  taken into account by a $z^{-n}$ block before the computation of controller
  reference input $\dot{\psi}_{\rm c}$.}
  \label{fig:block_ff_psidot}
\end{figure}

With such a feedforward/feedback structure we can accommodate the dependence of
the gain $K_{\dot{\psi}}$ from the air path velocity.
Fig.~\ref{fig:block_ff_psidot} provides details of the feedforward
controller block. In line with the idea of the model following principle the feedforward 
block is not limited to linear equations and can accommodate any system
description.
The extended version of the turn rate law (\ref{eq:Drehratengesetz_with_mass_term})
\begin{equation}\label{eq:equation1}
  \dot{\psi}_{\rm m} = K_{\dot{\psi}}\, \delta + M
  \frac{\cos\theta_{\rm g} \sin\psi_{\rm g}}{v_{\rm a}} 
\end{equation}
can easily be considered. The principal idea is to invert it in order to compute
the necessary deflection $\delta_{\rm ff}$.

In addition the block also contains nonlinear elements, like limiters on angle
and rate, in order to capture limited pod steering speed and other constraints.
That way the commands from the feedforward will never saturate the deflection
capability. Around $60\%$ of this range is reserved for the feedforward leaving
the remaining $40\%$ for the feedback. This is usually sufficient for the
feedback loop to counteract unmodeled plant uncertainties and disturbances.

The mass term  from (\ref{eq:equation1}) is introduced via
\begin{equation}
  T_1=\frac{M}{K_{\dot{\psi}}}\,\frac{\cos\theta_{\rm g} \sin\psi_{\rm
  g}}{v_{\rm a}}.\label{eq:T_1}
\end{equation}
\begin{figure}
  \centering
  \includegraphics[width=8.8cm]{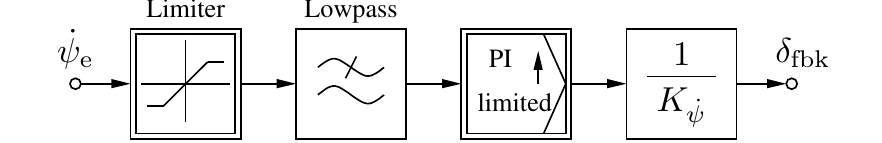}
  \caption{Feedback block $[{\rm C}_{\dot{\psi}}]$ of the cascaded controller (see
  Fig.~\ref{fig:cascaded_controller}). The controller mainly consists of
  a PI feedback on the yaw rate. Note the division by $K_{\dot{\psi}}$, which is 
  a function of the air path speed, thus
  introducing a nonlinearity into the feedback by transforming the rate
 command from the PI controller
  to the command portion $\delta_{\rm fbk}$ which is routed to the steering
  pod.}
  \label{fig:block_c_psidot}
\end{figure}
Fig.~\ref{fig:block_c_psidot} provides the details of the feedback controller
block:
As the plant has a proportional character the feedback structure is of proportional/integrator
(PI controller) type. The output of the PI controller is divided by $K_{\dot{\psi}}$ thus taking the
velocity dependence into account. 
A lowpass is added for noise rejection of the measurement.
\begin{figure}
  \centering
  \includegraphics[width=8.8cm]{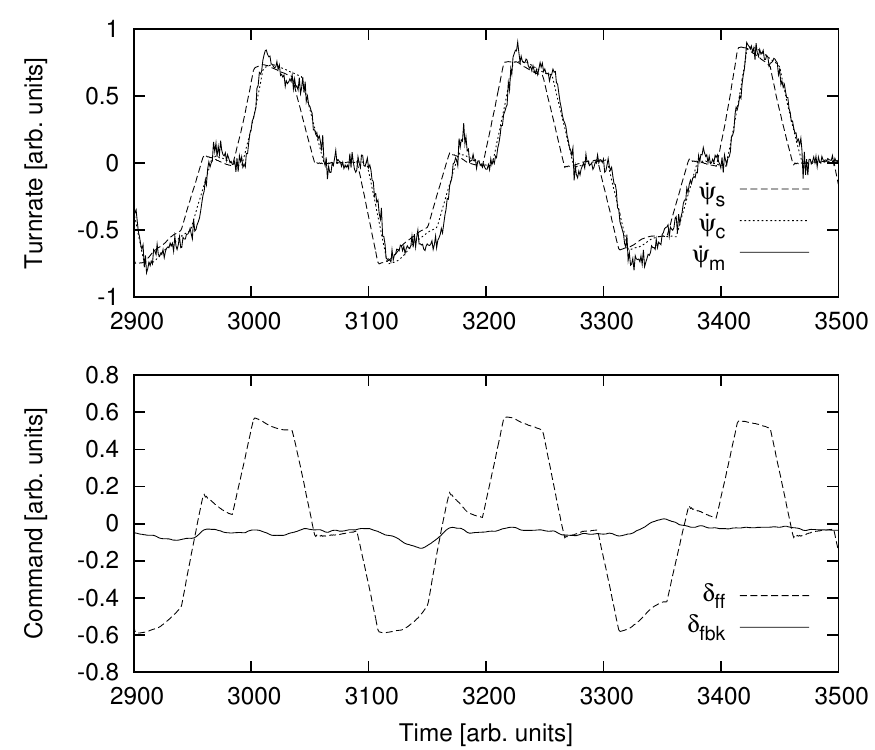}
  \caption{Experimental results for the inner loop. In the upper plot the
  measured yaw rate $\dot{\psi}_{\rm m}$ is compared to $\dot{\psi}_{\rm s}$ and
  $\dot{\psi}_{\rm c}$. The lower plot compares feedforward $\delta_{\rm ff}$ and
  feedback $\delta_{\rm fbk}$ controller output signals for a typical dynamical
  flight situation.}
  \label{fig:plot03}
\end{figure}

By providing Fig.~\ref{fig:plot03} we illustrate how effective the inner loop
actually works.
We show how a couple of repeating flight patterns look at the inner loop level. The feedforward 
command $\delta_{\rm ff}$ moves between $\pm 60\%$ of the available
deflection range as can be seen in the lower graph of the figure. The feedback command $\delta_{\rm fbk}$ hardly needs to correct control errors due
to unmodeled plant dynamics. Actually this figure is just an alternative
account to Fig.~\ref{fig:sysident2b} in proving the good fit of the turn rate
law.
\section{Controller Outer Loop}
\label{sec:controller_design_outer_loop}
\begin{figure}
  \centering
  \includegraphics[width=8.8cm]{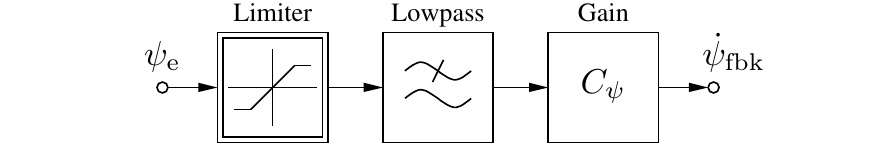}
  \caption{Feedback block $[{\rm C}_{\psi}]$ of the cascaded controller
  (compare Fig.~\ref{fig:cascaded_controller}). As the dynamics to be
  controlled is mainly an integrator, a proportional feedback has been chosen.}
  \label{fig:block_c_psi}
\end{figure}

\begin{figure}
  \centering
  \includegraphics[width=8.8cm]{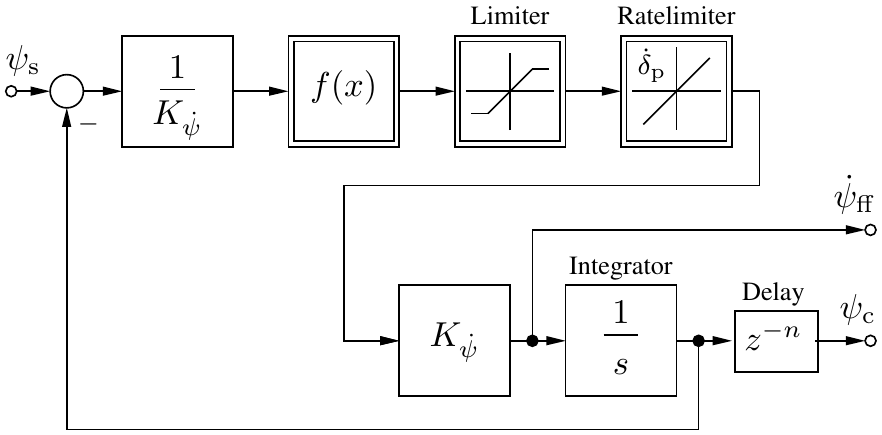}
  \caption{Feedforward block $[{\rm FF}_{\psi}]$ details of cascaded controller (see
  Fig.~\ref{fig:cascaded_controller}). The steering pod model is included as
  combination of limiter and ratelimiter. Note that even for step inputs on
  $\psi_{\rm s}$ the algorithm computes feedforward $\dot{\psi}_{\rm ff}$ and
  $\psi_{\rm c}$ in a way which is consistent with the capability of the overall dynamics. This is
  achieved by using an inner feedback loop embedding the steering pod model. 
Various delays in the whole loop are taken into account by a $z^{-n}$ block controller reference input $\psi_{\rm c}$.}
  \label{fig:block_ff_psi}
\end{figure}

As seen from the outer loop the inner loop has dealt with the aerodynamically influenced part
of the dynamics. It now remains for the outer loop controller to achieve a
desired ${\psi}_{\rm s}$. The division into feedforward
and feedback parts is kept.
Figs. \ref{fig:block_c_psi} and  \ref{fig:block_ff_psi} provide further
details.

\begin{figure}
  \centering
  \includegraphics[width=8.8cm]{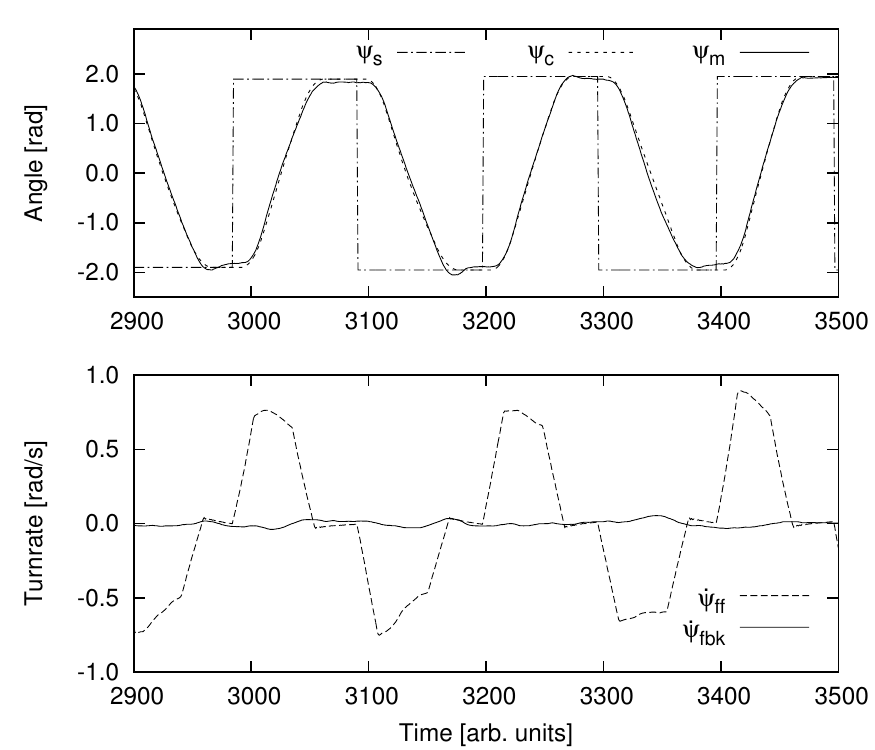}
  \caption{Flight results for the outer loop. The upper plot compares the
  response of the measured angle $\psi_{\rm m}$ to a given rectangular $\psi_{\rm s}$.
  The curve $\psi_{\rm c}$ is computed by the internal loop of the feedforward
  block $[{\rm FF}_{\psi}]$ using a model of the steering pod (compare
  Fig.~\ref{fig:block_ff_psi}).}
  \label{fig:plot04}
\end{figure}

As the plant characteristic is of integrating nature a feedback law  with
proportional character $[{\rm C}_\psi]$, augmented by a lowpass, is sufficient. As
before a limiter is also employed. The feedforward block $[{\rm FF}_{\psi}]$ (see
Fig.~\ref{fig:block_ff_psi}) has more elaborate features. Note that the
feedforward term features an internal feedback loop. The basic idea is to shape the
commanded $\psi_{\rm s}$, even if it is a jump, in such a way that it corresponds to
the actual response capabilities of the control pod and kite.
We achieve this by employing an internal loop from $\psi_{\rm s}$ to $\psi_{\rm c}$.
Note that the two limiters are crucial in shaping the final $\psi_{\rm c}$
evolution.
Furthermore we have a nonlinear function inside the loop:
$f(x)={\rm sign}(x)\sqrt{2\dot{\delta}_{\rm p} |x|}$. We will stop short in deriving
the details of this special feature and refer to Appendix \ref{sec:f_x}. 
Instead we will illustrate it with Fig.~\ref{fig:plot04}.
The upper part shows how the input command $\psi_{\rm s}$ is shaped into the
command $\psi_{\rm c}$ by the non-linear feature of the internal feedback loop.
$\psi_{\rm c}$ corresponds to an actually flyable  $\psi_{\rm c}$ pattern.
$\dot\psi_{\rm ff}$ shows the corresponding necessary rate pattern which is
fed into the inner loop.

As a conclusion we will summarize the basic design principles of our controller.
The first feature is the separation of the dynamics of deflection to rate
(\ref{eq:Drehratengesetz}) from the kinematic of rate to angle.
This separation allows us to introduce non-linear elements (mainly limiters) at
the appropriate places. This way we achieve a shaping of our commanded signals
such that they correspond to the limitations of the complete chain from software command over control pod
steering to kite movement.

A second characteristical feature is the feedforward/feed\-back separation which
allows us to decouple non-linear elements, as for example the mass term, from
the feedback. The feedback loops can then be designed within the realm of
linear control theory. Only proportional or integral dynamics remain which can be handled in a classical way. Due to the feedforward/feedback separation the selection of the closed
loop bandwidth of the two loops is more concerned with achieving sufficient stability margin than
with achieving performance in terms of fast response because this is already mastered by the feedforward.

The next chapter will illustrate the generation of the $\psi_{\rm s}$ command from
the desired kite trajectory. Although it could be perceived as just a further
cascaded loop we treat it more like the 'guidance' feature of classical
aerospace applications.
\section{Pattern Generation}
\label{sec:patterngeneration}
In this section we describe the dynamic flight mode which is
used to generate traction force by flying dynamical patterns in order to obtain
high air path speeds and forces. The algorithm utilizes the presented
controller design by providing the input value $\psi_{\rm s}$.

In order to explain the main principles we would like to review the
space station experiment of Section \ref{sec:basic_system}. In this model a
constant value of $\psi_{\rm s}=+\psi_0$ or $-\psi_0$ leads to a circular
orbit clockwise or counterclockwise dependent on the sign of $\psi_{\rm s}$ and the
obtained force can be easily controlled by the value of $\left|\psi_{\rm s}\right|$.
For the purpose of line force generation
in our space station experiment this would finish our design effort
--- but in our application we are not able to fly circular orbits as the kite
would crash onto the water surface. Thus the solution is to turn around at
certain points of the orbit and fly back and forth.

\begin{figure}
  \centering
  \includegraphics[width=8.8cm]{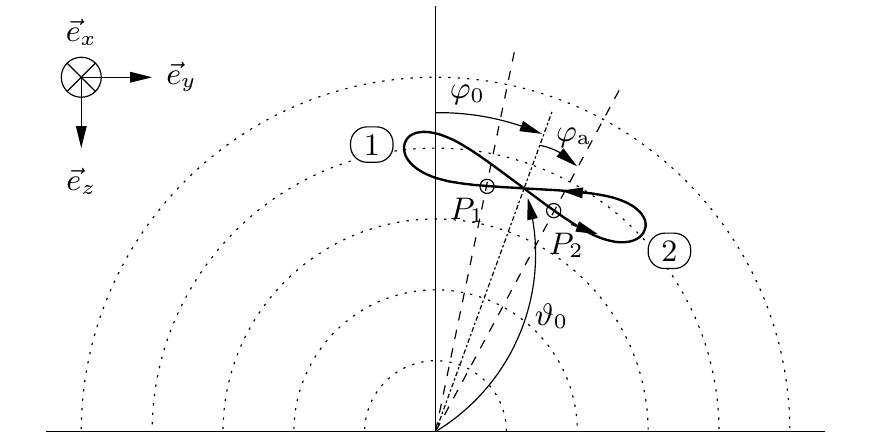}
  \caption{Geometry for pattern generation. The figure-eight pattern is
  guided by two states (1) and (2). Transitions between those are triggered
  by the conditions P$_1$: $\varphi \!<\! \varphi_0\!-\!\varphi_{\rm a}$ and
  P$_2$: $\varphi \!>\! \varphi_0\!+\!\varphi_{\rm a}$. The corresponding sequence is
  shown in Fig.~\ref{fig:pattern_sequence}. }
  \label{fig:pattern_geometry}
\end{figure}

\begin{figure}
  \centering
  \includegraphics[width=8.8cm]{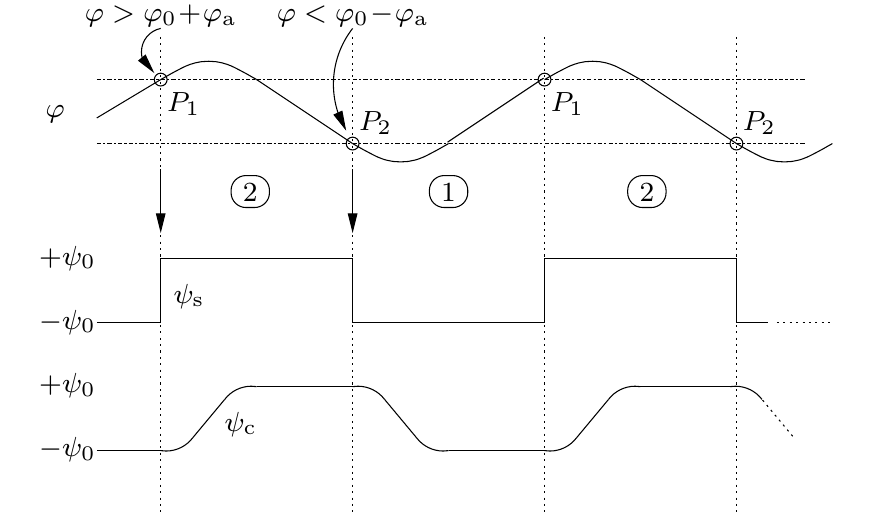}
  \caption{Pattern generation sequence toggling between the two states
  (1) and (2) when conditions P$_1$: $\varphi \!<\! \varphi_0\!-\!\varphi_{\rm a}$ and
  P$_2$: $\varphi \!>\! \varphi_0\!+\!\varphi_{\rm a}$ are met respectively. The
  states directly result in a square signal to $\psi_{\rm s}$ which is shaped by the
  feedforward block $[{\rm FF}_{\psi}]$ into $\psi_{\rm c}$ (compare
  Section \ref{sec:controller_design_outer_loop}). The respective pattern geometry
  is displayed in Fig.~\ref{fig:pattern_geometry}.}
  \label{fig:pattern_sequence}
\end{figure}

The resulting trajectory of such a scheme is shown in
Fig.~\ref{fig:pattern_geometry}.
The underlying algorithm is similar to those of the bang-bang experiments in
Section \ref{sec:turn_rate_law}. A constant value $\psi_{\rm s}=+\psi_0$ is
commanded until point 'P$_1$' is reached (at
$\varphi\!\leq\!\varphi_0\!-\!\varphi_{\rm a}$) triggering the command
$\psi_{\rm s}=-\psi_0$ until point 'P$_2$', then triggering the former value
$\psi_{\rm s}=+\psi_0$ and so on.
The timing of $\psi_{\rm s}$ is depicted in Fig.~\ref{fig:pattern_sequence}.

It is acceptable to apply a square signal to $\psi_{\rm s}$ as the controller
design contains an internal model which leads to a calculation of
$\psi_{\rm c}$. This calculation utilizes a given curve deflection and the given design
speed of the control pod (see Section \ref{sec:controller_design_outer_loop})
in order to perform the curve flight.
The determination of the optimal curve deflection or optimal turning
radius is involved as it depends on several geometric as well as
aerodynamic system characteristics \cite{Fagiano2009}, \cite{Fagiano2011} and goes beyond the scope of this paper.
For our system, the main effect during curve flight is the
decrease of air path speed due to the increase of $\vartheta$. 
Thus curve deflections are typically choosen in the range of 40--70\,\% of the
total deflection range in order to minimize the curve duration and optimize the
performance figure.

As illustrated in Fig.~\ref{fig:pattern_geometry} there are three
parameters defining the trajectory.
The parameter $\varphi_{\rm a}$ determines the
pattern size; the parameters $\varphi_0$ and $\vartheta_0$ determine the
center point of the pattern.
The value $\varphi_0$ can be freely chosen within
a certain range by the operator or an overlying algorithm in order to optimize the force
component pointing in forward direction of the vessel.
The value $\vartheta_0$ can not be tuned directly but indirectly by the
$\psi_0$ value which is the tuning knob for the air path speed $v_{\rm a}$ and hence
force.
Nevertheless (\ref{eq:v_a_theta}) does not hold in its simple
way and could by improved by using $\left<\psi\right>=\int dt
\left|\psi(t)\right|$ instead of $\psi_0$ in order to estimate $\vartheta$
or resulting forces.

\begin{figure}
  \centering
  \includegraphics[width=8.8cm]{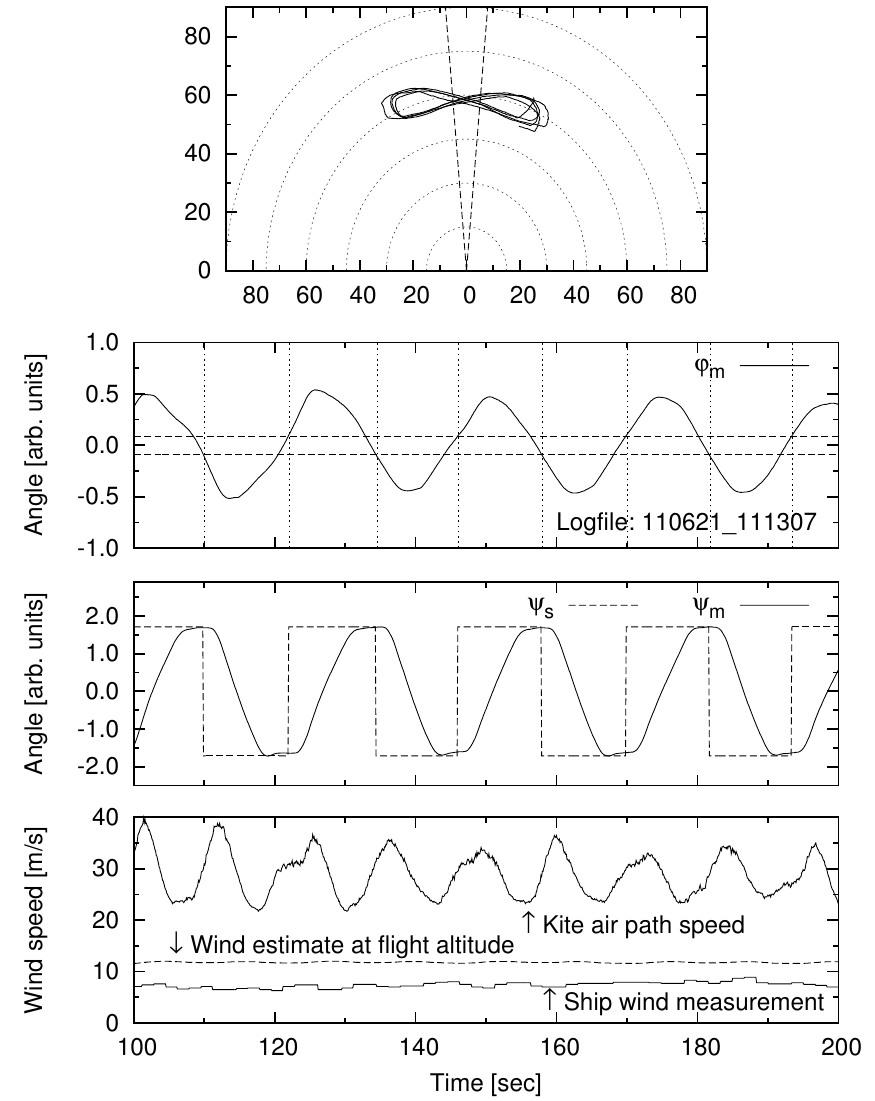}
  \caption{Flight test results illustrating pattern generation using a
  160\,m$^2$ kite at an operational towing line length of 300\,m.
  The trajectory for a typical figure-eight pattern is shown in the upper plot
  (compare Fig.~\ref{fig:pattern_geometry}).
  The two plots in the middle show corresponding curves for $\varphi_{\rm m}$,
  $\psi_{\rm c}$ and $\psi_{\rm m}$ (see Fig.~\ref{fig:pattern_sequence}).
  The lower data plot shows the measured wind speeds. Note that the dynamical flight mode
  leads to an air path speed of factor 3--4 higher than the
  estimated wind speed at flight altitude which is significantly higher than the wind speed measured
  aboard the ship.}
  \label{fig:plot06}
\end{figure}

As it is cumbersome to predict the exact wind situation at flight altitude in
any case we make use of another approach for force control: we use an outer loop
force controller evaluating the height of the force peaks while flying
the figure-eight pattern which provides a feedback value for $\psi_0$.
Details of this control law as well as of the supervision mechanisms during operative flight
and start procedure of the pattern are important and interesting issues each
but would exceed the purpose of this paper.
In Fig.~\ref{fig:plot06} we present the trajectory of some eights and show the
corresponding time series of the angles $\varphi_{\rm m}$, $\psi_{\rm s}$, $\psi_{\rm c}$ as
well as of the air path and wind speeds.

For sake of completeness we would like to explain the control
strategies for the neutral flight mode introduced in
Section \ref{sec:basic_system}.
The determining equation is (\ref{eq:dot_phi}) and the issue is to control
$\varphi$ to the given set value $\varphi_0$. We mainly use a linear
controller based on a classical PID controller in order to compute
$\psi_{\rm s}$ from $\varphi_{\rm m}\!-\!\varphi_0$. 
\section{Discussion and Further Challenges}
Dealing with a complex system in a demanding environment we could
present several further topics on our control system which are closely related
to the discussed contents.
These topics have been omitted in the
previous sections because we wanted a clear outline of the major features
of the algorithm. They are now briefly summarized to give a more
comprehensive understanding.

First of all we have completely excluded the ship from our treatment by 
arguing that consideration of the apparent wind, which is the wind speed with
respect to the ship's motional frame of reference, is an appropriate
approximation for the basic design of the control system. Furthermore the
choice of pattern parameters \cite{Dadd2011}, \cite{Fagiano2011} the
optimization of the towing force with respect to the ship forward direction 
\cite{Houska2006}, \cite{Fagiano2012} as well as consideration of influences on
dynamics due to waves are improvements to an operational kite autopilot but
have not been presented here.

It is worth mentioning that both, the thorough choice of the sensor set up and
preprocessing algorithms, contribute a significant and crucial part to the
presented controller performance and robustness. 
An important task is the estimation of the angle $\psi$ which involves
not only sophisticated inertial navigation algorithms but also takes
into account estimates for wind speed and wind direction at flight
altitude as these may vary on the timescale of minutes. 
A further discussion of these mainly technical and cumbersome topics will be
subject to future publications.

In our discussions we have made use of constant and long line lengths which is
the common operating point of our system.
Winching of the towing line is done while launching and recovering our
system and goes along with extra dynamic effects which are considered as
additional correction terms to the presented equations.
Especially for control during launch and recovery at shorter line lengths we use
the idea of receding horizon feedback from the model predictive control (MPC)
approaches \cite{JMaciejowski2002} in order to provide $\psi_{\rm s}$. 
However our optimization computation is done analytically
(thanks to the special and simple structure of the design model, see \ref{sec:system_dynamics})
as opposed to having a numerical solver. 

Wind gusts may also go along with downwinds leading to a free flight situation
of the usually constrained system. This situation involves a completely
different dynamics. While (\ref{eq:Drehratengesetz}) still
holds in large part the kinematic changes significantly. It is a crucial
advantage of the particular choice of coordinate system (see section
\ref{sec:basic_system}) that controller input values behave in a favorable way during these exceptional
occurrences. We would like to note that also disturbances due to
excitations of internal modes of the real system are effectively suppressed 
although these modes are not considered explicitly in the 'design model'.

A more elaborate challenge is the avoidance of and response to stall situations.
Systematic experimental tests on this are hardly feasible. Nevertheless it is
an important issue and subject to current research and development activities.
\section{Summary}
In this paper we have presented a simple dynamical model for the dynamics
of constrained kites. A key point has been that for controller design the
complete aerodynamics can be reduced to a law involving only
one   or two
quantities (see (\ref{eq:Drehratengesetz}) 
and  (\ref{eq:Drehratengesetz_with_mass_term}), respectively).
We have discussed flight data to justify that this reduction describes
reality to a surprisingly high degree. Utilizing this we have developed a
cascaded controller, based on the model following principle approach,
and proved the effectiveness with flight test data. Finally
the main principles of flight pattern generation have been given.

We would like to finish this paper by emphasizing that the presented
results hold for the SkySails towing kite system but the underlying equations
are commonly valid for the dynamics of constrained kites. Thus a lot more
applications of the presented controller are feasible especially in the
fascinating upcoming field using kites in order to generate electricity and
to further open up the green resource of wind power in higher altitudes and
off-shore.
\appendices
\section{Derivation of System Dynamics}
\label{sec:derivation_model}
The system vectors read (see Fig.~\ref{fig:coordinate_definition}):
\begin{IEEEeqnarray}{ccc}
  \vec{e}_{\rm roll} &=&
     \left(\begin{array}{c}-\sin\vartheta \cos\psi \\
     -\cos\varphi\sin\psi + \sin\varphi \cos\vartheta \cos\psi\\
     -\sin\varphi \sin\psi -\cos\varphi \cos\vartheta
     \cos\psi\end{array}\right)\\
  \vec{e}_{\rm pitch} &=& \left(\begin{array}{c}\sin\vartheta \sin\psi \\
-\cos\varphi\cos\psi -\sin\varphi \cos\vartheta \sin\psi\\
-\sin\varphi \cos\psi + \cos\varphi\cos\vartheta \sin\psi
\end{array}\right)\\
   \vec{e}_{\rm yaw} &=& \left(\begin{array}{c}-\cos\vartheta \\ -\sin\varphi
   \sin\vartheta \\ \cos{\varphi}\sin\vartheta \end{array}\right).
\end{IEEEeqnarray}
As discussed in Section \ref{sec:basic_system} we neglect gravity. Thus the
problem becomes independent of $\varphi$ and can be treated for~$\varphi=0$
without loss of generality which implies the following
basis vectors:
\begin{IEEEeqnarray}{ccc}
   \vec{e}_{\rm roll} &=& \left(\begin{array}{c}-\sin\vartheta \cos\psi \\ -\sin\psi\\
  -\cos\vartheta \cos\psi\end{array}\right)\label{eq:e_r}\\
  \vec{e}_{\rm pitch} &=& \left(\begin{array}{c}\sin\vartheta \sin\psi \\ -\cos\psi \\ \cos\vartheta \sin\psi \end{array}\right)\\
  \vec{e}_{\rm yaw} &=& \left(\begin{array}{c}-\cos\vartheta \\ 0 \\ \sin\vartheta
  \end{array}\right).\label{eq:e_y}
\end{IEEEeqnarray}
The air flow $\vec{v}_{\rm a}$ of the flying system is
\begin{equation}
  \vec{v}_{a} = \left(\begin{array}{c} v_0 \\ 0 \\ 0\end{array}\right) - v_{\rm roll}
  \vec{e}_{\rm roll} - v_{\rm pitch} \vec{e}_{\rm pitch}.\label{eq:v_a}
\end{equation}
The first term describes the external wind, the subsequent two terms the
flow due to the kinematic speeds $v_{\rm roll}$ and $v_{\rm pitch}$ with respect to the
basis vectors $\vec{e}_{\rm roll}$ and $\vec{e}_{\rm pitch}$.

Considering the basic aerodynamics of an
airfoil \cite{JLingard1986} as a very simple model we claim validity of
the following two conditions:
\begin{enumerate}
\item The airflow vector lies in the 
$\vec{e}_{\rm roll}$-$\vec{e}_{\rm yaw}$-plane which means
  \begin{equation}
    \left(\vec{e}_{\rm pitch},\vec{v}_{\rm a}\right) = 0.\label{eq:flightcondition01}
  \end{equation}
\item The airflow direction with respect to $\vec{e}_{\rm roll}$ and
$\vec{e}_{\rm yaw}$ is given by the glide ratio $E$ which is the ratio between lift
and drag coefficients $E=L/D$,
  \begin{equation}
    \frac{\left(\vec{e}_{\rm roll},\vec{v}_{\rm a}\right)}{\left(\vec{e}_{\rm yaw},\vec{v}_{\rm a}\right)} =
    E.\label{eq:flightcondition02}
  \end{equation}
\end{enumerate}
Insertion of the definitions (\ref{eq:e_r})--(\ref{eq:e_y}) into
(\ref{eq:flightcondition01}) and (\ref{eq:flightcondition02}) yields the
velocity components
\begin{eqnarray}
  \frac{v_{\rm pitch}}{v_0} &=& \sin\vartheta \sin\psi \label{eq:result_v_p}\\
  \frac{v_{\rm roll}}{v_0} &=& E \cos\vartheta - \sin\vartheta
  \cos\psi.\label{eq:result_v_r}
\end{eqnarray}
The airflow in roll direction $v_{\rm a}$, measured by the anemometer in the
control pod, can be calculated using (\ref{eq:e_r}) and (\ref{eq:v_a}) by:
\begin{equation}
  v_{\rm a} = -\left(\vec{v}_{\rm a}, \vec{e}_{\rm roll}\right) = v_0 E
  \cos\vartheta.\label{eq:v_g}
\end{equation}
Geometric and kinematic considerations lead to the relation
\begin{equation}
  \dot{\vartheta} = \frac{1}{L} \left(v_{\rm roll} \cos\psi - v_{\rm pitch}
  \sin\psi\right).
\end{equation}
Using (\ref{eq:result_v_p}), (\ref{eq:result_v_r}) and (\ref{eq:v_g}) we get for
the dynamics of $\vartheta$ the following differential equation:
\begin{equation}
  \dot{\vartheta} = \frac{v_{\rm a}}{L} \left(
  \cos\psi - \frac{\tan\vartheta}{E}\right)
  .\label{eq:vartheta_dot}
\end{equation}
Similarly we obtain for $\varphi$ the equation
\begin{equation}
  \dot{\varphi} = \frac{1}{L\sin\vartheta}\left(-v_{\rm roll} \sin\psi - v_{\rm pitch} \cos\psi
  \right),
\end{equation}
and using (\ref{eq:result_v_p}), (\ref{eq:result_v_r}) and (\ref{eq:v_g}) the
equation of motion
\begin{equation}
  \dot{\varphi} = -\frac{v_{\rm a}}{L \sin\vartheta} \sin\psi. 
  \label{eq:varphi_dot}
\end{equation}
For constant values of $\psi$, one obtains the steady-state solution of
(\ref{eq:vartheta_dot}) as
\begin{equation}
  \vartheta_0(\psi) = \arctan(E \cos\psi).
\end{equation}
\section{Equations of Motion}
\label{sec:eqm_summary}
In this appendix we summarize the equations of motion
(\ref{eq:Drehratengesetz}), (\ref{eq:vartheta_dot}) and (\ref{eq:varphi_dot})
for our model
\begin{eqnarray}
  \dot{\psi} &=& g\, v_{\rm a}\, \delta \label{eq:eqm11}\\  
  \dot{\vartheta} &=& \frac{v_{\rm a}}{L} \left(
  \cos\psi - \frac{\tan\vartheta}{E}\right)\\
  \dot{\varphi} &=& -\frac{v_{\rm a}}{L \sin\vartheta} \sin\psi.\label{eq:eqm13}
\end{eqnarray}
Following our model we find that $v_{\rm a}$ is a function of the ambient wind
speed $v_0$ and $\vartheta$ (\ref{eq:v_g}) and has to be considered as part of
the equations of motions:
\begin{equation}
  v_{\rm a} = v_0 E\cos\vartheta.\label{eq:v_g2}
\end{equation}
Before inserting this relation into (\ref{eq:eqm11})--(\ref{eq:eqm13}), we would
like to emphasize that $v_{\rm a}$ can be measured directly as a single sensor
input instead of using (\ref{eq:v_g2}).
Using $v_0$ and $\vartheta$ to determine $v_{\rm a}$ would introduce avoidable
inaccuracies into our control loop as in addition to errors in the aerodynamical
model and $E$ the wind speed at flight altitude, which should by used for
$v_0$, is typically (but not necessarily) higher than the wind speed measured
aboard the vessel (compare Fig.~\ref{fig:plot06}). Computing an
estimate for $v_0$ at flight altitude involves $v_{\rm a}$ and thus would not
provide any benefit compared to using $v_{\rm a}$ directly. In other words
(\ref{eq:eqm11}) represents the physics of the kite reacting to a deflection
$\delta$ with a turn rate $\dot{\psi}$ scaled by the airflow speed $v_{\rm a}$
independent of e.g.~the position represented by $\vartheta$.

In contrast to an operational control setup a numerical simulation has to
compute $v_{\rm a}$ based on $v_0$. The set of equations can be combined by
inserting (\ref{eq:v_g2}) into (\ref{eq:eqm11})--(\ref{eq:eqm13}) and we obtain:
\begin{eqnarray}
  \dot{\psi} &=& (g\, v_0 E \cos\vartheta ) \, \delta \\  
  \dot{\vartheta} &=& \frac{v_0}{L} \left(E
  \cos\vartheta \cos\psi - \sin\vartheta\right)\\
  \dot{\varphi} &=& -\frac{v_0\, E}{L \tan\vartheta} \sin\psi.
\end{eqnarray}
\section{Function $f(x)$ for $\psi$-Feedforward Block}
\label{sec:f_x}
In this appendix the origin of the
nonlinear function $f(x)$ used by the feedforward block in
Section \ref{sec:controller_design_outer_loop} is briefly outlined.
We thereby emphasize that the following is not crucial for an
understanding of the main concepts but gives further insight
into a model-based detail of the feedforward generation.
Assume the process of starting with an
initial deflection of $\delta\!=\!\delta_{\rm i}$ and steering to $\delta\!=\!0$
at a constant velocity $\dot{\delta}_{\rm p}$ .
The corresponding change in $\psi$ which we denote by $\Delta\psi$ can be
computed using (\ref{eq:Drehratengesetz}) as
$\Delta\psi \!=\! \int dt \dot{\psi} \!=\! K_{\dot{\psi}}\int
dt\,\dot{\delta}_{\rm p} t$ for $t=0..(\delta_{\rm i}/\dot{\delta}_{\rm p})$. Resolving with respect to $\delta_{\rm
i}$ yields:
\begin{equation}
  \delta_{\rm i} = {\rm sign}(\Delta\psi) \sqrt{\frac{2 \dot{\delta}_{\rm p}
  |\Delta\psi|}{K_{\dot{\psi}}}}.\label{eq:psi_to_delta}
\end{equation}
This solution already includes the bookkeeping of signs assuming
that $\dot{\delta}_{\rm p}$ is a positive parameter. 
The interpretation of the result is as follows: for a given
deviation $\Delta\psi$ of the internal state to the set point $\psi_{\rm s}$
(compare Fig.~\ref{fig:block_ff_psi}) a $\delta_{\rm i}$ can be calculated using
(\ref{eq:psi_to_delta}). This $\delta_{\rm i}$ describes the maximum deflection
allowed for the internal model so that no overshoots occurs
when the internal feedback loop reduces the error $\Delta\psi$ under the
assumption of constant $\psi_{\rm s}$ and $K_{\dot{\psi}}$. 
Thus $\delta_{\rm i}$ is a suitable input for the model shaping
elements limiter and ratelimiter in Fig.~\ref{fig:block_ff_psi}.
Finally $f(x)={\rm sign}(x)\sqrt{2\dot{\delta}_{\rm p}|x|}$ can be deduced from
comparing (\ref{eq:psi_to_delta}) to Fig.~\ref{fig:block_ff_psi}.
\section*{Acknowledgements}
\addcontentsline{toc}{section}{Acknowledgments}
We acknowledge support by the whole SkySails team,
especially contributions from the kite, software, hardware and mechanical
development teams.
Their knowledge and excellent work on the system as well as the unfatiguing 
support by the test engineers and nautical crews during numerous sea trials are
crucial contributions to the findings presented here.
%
%

%
%
\vfill 
\pagebreak
\begin{IEEEbiographynophoto}{Michael Erhard} received the Diploma degree
from the University of Freiburg, Freiburg, Germany, and the Ph.D.~degree, which
involved research on multi-component Bose--Einstein
condensates, from the University of Hamburg, Hamburg, Germany, in 2000 and
2004, respectively, both in physics. His
research interests included experiments in quantum optics, laser physics, and
theoretical quantum optics.

He joined SkySails GmbH, Hamburg, Germany, in 2004, as a Development Engineer,
where he was deeply involved in several hardware and software designs
on the sensor data acquisition system and the kite steering units. 
He is currently in charge of the flight control system (autopilot). His main
responsibility is the development of sensor data processing and control algorithms for
the flying system and evaluation of those algorithms in flight tests.
\end{IEEEbiographynophoto}%
%
%
%
\begin{IEEEbiographynophoto}{Hans Strauch} received the Diploma degree
in physics (1981) from the University of Kiel, Kiel, Germany.

He joined Ansch\"utz GmbH, Kiel, where he was involved in the development of
autopilots and ground track controllers for ships. In 1988, he 
joined Astrium Space Transportation, Bremen, Germany, where he was
involved in development of the guidance and control algorithms for various
vehicles ranging from upper stages of launchers to re-entry bodies. 
He is currently holds a Senior Expert GNC.
From 1998 to 2002, he was with NASA, Johnson Space Center, Houston, TX, where he
was involved in the development of guidance and control software for the
parafoil phase of the X38 Crew Rescue Vehicle.
In 2004, he was responsible for the attitude control of the winged body
autonomous landing demonstrator PHOENIX.
He has been a Consultant with SkySails GmbH, Hamburg, Germany since 2004. 
His contributions to the findings reported in this paper are independent from
his affiliation to Astrium.
\end{IEEEbiographynophoto}%
%
\end{document}